\documentclass[12pt]{article}
\usepackage{graphicx}
\usepackage{amsfonts}
\usepackage{amssymb,amsmath}
\usepackage{latexsym}
\usepackage{color}
\usepackage{cite}
\input{colordvi.tex}

\begin{document}

\begin{titlepage}

\begin{flushright}
IPMU 09-0133\\
ICRR-Report-554-2009-16
\end{flushright}

\begin{center}

{\Large \bf  
Probing the primordial power spectra with inflationary priors
}

\vskip .45in

{\large
Masahiro Kawasaki$^{1,2}$,
Toyokazu Sekiguchi$^1$}

\vskip .45in

{\em
$^1$Institute for Cosmic Ray Research, University of Tokyo,
Kashiwa 277-8582, Japan  \vspace{0.2cm} \\
$^2$Institute for the Physics and Mathematics of the Universe,
University of Tokyo, Kashiwa, Chiba, 277-8568, Japan \vspace{0.2cm}\\
}

\end{center}

\vskip .4in

\begin{abstract}
We investigate constraints on power spectra of the primordial curvature and 
tensor perturbations with priors based on single-field slow-roll inflation models. 
The Hubble slow-roll parameters are included in cosmological parameters
and the primordial power spectra are generated using the inflationary flow equations.
Using data from recent observations of CMB and several 
measurements of geometrical distances in the late Universe, 
we perform Bayesian parameter estimation and model selection 
for models that have separate priors on the 
slow-roll parameters. The same analysis is also performed adopting 
the standard parameterization of the primordial power spectra.
We confirmed that the scale-invariant Harrison-Zel'dovich spectrum is disfavored
with more significance than previous studies.
While current observations appear to be optimally modeled with 
some simple models of single-field slow-roll inflation, 
data is not enough constraining to distinguish these models.
\end{abstract}

\end{titlepage}

\setcounter{page}{1}

\section{Introduction} 
\label{sec:introduction}
Cosmological observations, including the cosmic microwave background 
(CMB), large scale structure, baryon acoustic oscillation (BAO), type Ia 
supernovae (SN) offer opportunities to probe new
physics far beyond the reach of experiments in terrestrial laboratories. One of such
physics is inflation, that solves various problems in the hot universe scenario 
\cite{Starobinsky:1980te,Guth:1980zm,Sato:1980yn,Linde:1981mu,Albrecht:1982wi}.
In addition, inflation also explains generation of initial perturbations for structure 
formation~\cite{Starobinsky:1979ty,Mukhanov:1981xt,Hawking:1982cz,Starobinsky:1982ee,
Guth:1982ec,Linde:1982uu,Bardeen:1983qw,Abbott:1984fp}. 
At present, Inflation is an essential part of our best description of the Universe.

The simplest class of models of inflation is so-called single-field slow-roll inflation
\cite{Linde:1981mu,Albrecht:1982wi}, where potential energy of 
a single scalar field (inflaton), whose field value varies slowly, drives 
an exponential expansion of the Universe. Gaussian, nearly scale-invariant primordial 
curvature perturbation can be generated from the vacuum fluctuation of inflaton in a 
quasi-de Sitter spacetime. Such primordial perturbation gives excellent fits to 
various data from cosmological observations, which makes single-field slow-roll inflation
highly attractive. However we yet know quite little about inflaton and its potential.

Identification of inflaton is of particular interest both in cosmology and particle physics. 
As a model of single-field slow-roll inflation left its vestige in the late Universe in an 
observable way through generation of the primordial curvature 
and tensor perturbations, constraints on models have been mainly 
investigated through constraining power spectra of these primordial perturbations, $\mathcal P_\zeta(k)$ and
$\mathcal P_h(k)$, using various cosmological observations\footnote{
There are also several other probes for inflation models, including 
primordial isocurvature perturbations and 
non-Gaussianity in primordial perturbations.
}. One of the most familiar ways may be to adopt the standard parameterization for the power spectra, 
$A_s$, $r$, $n_s$, $n_t$, $\alpha_s$, etc. (See Eqs.~(\ref{eq:A_s}-\ref{eq:n_t})),
and derive constraints on these parameters. Some more involved analyses 
have also been performed focusing on reconstruction of the potential of the inflaton or slow-roll 
flow parameters \cite{Copeland:1993jj,Lidsey:1995np,Leach:2002dw,
Easther:2002rw,Kinney:2003uw,Leach:2003us,Peiris:2006ug,
Kinney:2006qm,Martin:2006rs,Peiris:2006sj,Finelli:2006fi,Lesgourgues:2007gp,
Powell:2007gu,Lesgourgues:2007aa,Bean:2008ga,Hamann:2008pb,
Adshead:2008vn,Agarwal:2008ah,Powell:2008bi}. In either way, deviation from 
the Harrison-Zel'dovich (HZ) spectrum (i.e. $r=0$ and $n_s=\alpha_s=...=0$) 
allows us to probe models of single-field slow-roll inflation. Cosmological data
now signifies some spectral features in the power spectrum of the curvature perturbation.
When we assume a power-law for curvature perturbation spectrum $\mathcal P_\zeta(k)\propto A_s k^{n_s-1}$ 
and absence of the tensor perturbation $\mathcal P_h(k)=0$, 
recent CMB data from WMAP give $-0.065<n_s-1<-0.009$ (95\% C.L.), suggesting
a significant deviation from the HZ power spectrum \cite{Komatsu:2008hk}. 

However, constraints on the primordial power spectra
are highly dependent on parameter spaces which we investigate.
For instance, if running of the spectral index $\alpha_s$ is included, 
the same data allow relatively large $\alpha_s$ \cite{Komatsu:2008hk}.
This kind of issues commonly arise when we try to constrain
the primordial power spectra in parametric ways.
Consequently, shapes of reconstructed power spectra also
differ depending on the choice of parameter spaces, which
finally affects selection among inflation models.

Since we do not in advance know a parameter space where we should explore 
a constraint on the power spectrum, we also need to examine whether the parameter space is
appropriate. More generally, appropriateness of a model, which possesses its own prior 
assumption, should also be discussed.
A guiding principle in looking for an optimal model is Occam's razor, which
penalizes unnecessary assumption in describing observations.
Bayesian model selection is Bayesian implementation of Occam's razor, 
which is now frequently applied in the context of 
cosmology~\cite{Slosar:2002dc,Beltran:2005xd,
Trotta:2005ar,Bridges:2005br,Kunz:2006mc,Parkinson:2006ku,Bridges:2006zm,
Pahud:2006kv,Liddle:2006tc,Heavens:2007ka,Gordon:2007xm,
Mukherjee:2008pq,Trotta:2008qt,Mukherjee:2008zzb,Bridges:2008ta,
Liddle:2009xe,Sollom:2009vd,Ichikawa:2009ir,Valiviita:2009bp}.
In particular, the authors in Ref. \cite{Lesgourgues:2007gp} have adopted Bayesian model selection
to assess optimal orders up to which reconstruction of inflaton potential should be performed. 
Moreover, in Ref. \cite{Ballesteros:2007te} Bayesian model selection is directly applied
to distinguish some class of inflation models.

Motivated by \cite{Lesgourgues:2007gp,Ballesteros:2007te} 
and other earlier studies, in this paper we investigate an optimal constraint 
on primordial perturbation spectra and make comparison 
of single-field slow-roll inflation models using recent cosmological observations, 
which would be a subject of great interest~\cite{Baumann:2008aq}. 
For these purpose we make vigorous use of Bayesian model selection.
We compare Bayes evidences for several models which have
separate priors on inflationary slow-roll parameters, or parameters of primordial power spectra.
Each of these models can be regarded as representing some class of single-field slow-roll inflation models.
In this paper we implicitly assume a Freedman-Robertson-Walker universe and adopt natural units
$\hbar=c=M_\mathrm{Pl}=1$.

This paper is organized as follows: In Section~\ref{sec:HSR} we briefly review 
the Hubble slow-roll flow equations, which are adopted in our analysis.
Then in Section~\ref{sec:selection}, some essences of Bayesian model selection are also reviewed. 
The main part of our paper is Section~\ref{sec:analysis}, where we investigate
constraints on the primordial power spectra using data from recent observations 
of CMB and geometrical distances in the late Universe. By employing
Bayesian model selection, an optimal constraint and comparison of models 
of single-field slow-roll inflation are investigated. The final section is devoted to 
summary and future outlook.

\section{Hubble slow-roll flow equations}
\label{sec:HSR}
We make use of the {\it Hubble slow-roll} (HSR) flow equations~\cite{Leach:2002ar,
Leach:2002dw,Leach:2003us}, a specific version of 
more general inflationary flow equations \cite{Hoffman:2000ue,Kinney:2002qn}. 
Our algorithm is basically identical to slow-roll reconstruction presented 
in \cite{Peiris:2006sj,Adshead:2008vn} and we here quote some key consequences. 
In this paper, we assume models of single-field slow-roll inflation have canonical 
kinetic terms and do not investigate non-canonical models.

During an epoch of slow-roll inflation, the inflaton field $\phi$ can be safely 
assumed as a monotonic function of the time and hence regarded as a 
generalized time coordinate. Instead of solving the Hamilton-Jacobi 
equation~\cite{Kinney:2002qn} for given inflaton potential, 
we rather solve a set of the HSR flow equations given by
\begin{eqnarray}
\epsilon_{H}(\phi)&=&\frac{1}{4\pi}\left(\frac{H(\phi)^\prime}{H(\phi)}\right)^2, \label{eq:flow0} \\
^\ell\lambda_H(\phi)&=&\left(\frac{1}{4\pi}\right)^\ell
\frac{(H(\phi)^\prime)^{\ell-1}}{H(\phi)^\ell}\frac{d^{\ell+1}H(\phi)}{d\phi^{\ell+1}} 
\hspace{5mm}(\mbox{for $\ell\ge1$}). \label{eq:flow1}
\end{eqnarray}
Here and hereafter we denote derivatives respective to $\phi$
as prime (e.g. $ H^\prime\equiv dH/d\phi$).
Notice that $\phi$ plays a role of time variable in this formalism.
These flow equations are solved once we specify the initial values for 
$\epsilon_{H}$ and $^\ell\lambda_H$ at a fiducial $\phi_*$, 
\begin{eqnarray}
\epsilon_{H*}&\equiv& \epsilon_H(\phi_*), \\
^\ell\lambda_{H*}&\equiv&^\ell\lambda_H(\phi_*)\hspace{5mm}(\mbox{for $\ell\ge1$}).
\end{eqnarray}
If we choose fiducial HSR parameters $^\ell\lambda_{H*}=0$ for $\ell>M$,
the HSR parameters $^\ell\lambda_H(\phi)$ for $\ell>M$ also vanish at any 
$\phi$ and hence the flow equations are truncated at order $M$.
In this case, the Hubble expansion rate can be exactly solved~\cite{Liddle:2003py}.
Without loss of generality, we can choose $\phi_*=0$ and $\phi$ being a 
decreasing function of time (i.e. $H^\prime(\phi)>0$). Then we obtain
\begin{eqnarray}
H(\phi)=H_*(1+B_1\phi+B_2\phi^2+\cdots B_{M+1}\phi^{M+1}), \label{eq:Hubble}
\end{eqnarray}
where $H_*=H(\phi_*)=H(0)$ and the coefficients of the Taylor expansion, $B_i$ 
($i=1, \dots, M+1$), are given by the HSR parameters at the fiducial point,  
\begin{eqnarray}
B_1&=&\sqrt{4\pi\,\epsilon_{H*}}, \\
B_{\ell+1}&=&\frac{(4\pi)^\ell}{(\ell+1)!B_1^{\ell-1}}\,^\ell\lambda_{H*}\hspace{5mm}(\mbox{for $\ell\ge1$}).
\end{eqnarray}
Since we have specified the function $H(\phi)$, we already know the dynamics of 
the background universe and related variables as functions of $\phi$. For example, 
e-folding number $N(\phi)$ at time $\phi$ is given by integrating
\begin{eqnarray}
\frac{dN}{d\phi}=-\sqrt{\frac{4\pi}{\epsilon_H(\phi)}}.
\end{eqnarray}
Similarly, a wave number that exits the horizon at time $\phi$ is given 
by integrating
\begin{eqnarray}
\frac{d\ln k}{d\phi}=-\sqrt{\frac{4\pi}{\epsilon_H(\phi)}}(1-\epsilon_H(\phi)).
\label{eq:wnumber}
\end{eqnarray}
Once we fix a fiducial wave number $k_*=k(\phi_*)$, there is a one-to-one 
correspondence in $k$ and $\phi$ by Eq.~(\ref{eq:wnumber}), as long as 
slow roll of the inflaton does not break down. Then any other function of $\phi$ 
is rewritten as that of $k$. For later convenience, we choose the e-folding number 
$N(k)$ so that it vanishes at a wave number $k=10^{-4}$Mpc$^{-1}$, which 
roughly corresponds to the scale of current horizon $\simeq10$ Gpc.

In this paper, we truncate the HSR flow equations at $M=2$. This is because 
higher order HSR parameters may not be important due to the limited range of 
observable wave numbers 
$\mathcal{O}(10^{-4})<k<\mathcal{O}(0.1)~\mbox{Mpc}^{-1}$ and insufficient accuracy of 
data at present or in the near future. The fiducial HSR parameters up to $M=2$ 
are nothing but the usual slow-roll parameters $\epsilon(\phi)$, $\eta(\phi)$ and 
$\xi(\phi)$ and they are given by
\begin{eqnarray}
\epsilon(\phi)&=&\epsilon_H(\phi),\\
\eta(\phi)&=&\,^1\lambda_{H}(\phi), \\
\xi(\phi)&=&\,^2\lambda_{H}(\phi).
\end{eqnarray}

One of the most important prediction of single-field inflation models is that the 
primordial curvature and tensor perturbations are not independent of but related 
to each other since both are generated from the dynamics of the single scalar field 
$\phi$. Thus there exists the consistency relation for single-field 
inflation models. At second order in slow-roll approximation, the power spectrum 
of the primordial curvature and tensor perturbations, $\mathcal{P}_\zeta(k)$ and 
$\mathcal{P}_h(k)$, respectively, are given by~\cite{Stewart:1993bc}
\begin{eqnarray}
\mathcal{P}_\zeta(k)&=&\left.\frac{[1-(2C+1)\epsilon(\phi)
+C\eta(\phi)]^2}{\pi\epsilon(\phi)}H(\phi)^2\right|_{\phi=\phi(k)}, \label{eq:scal}\\
\mathcal{P}_h(k)&=&\left.\frac{16[1-(C+1)\epsilon(\phi)]^2}{\pi}
H(\phi)^2\right|_{\phi=\phi(k)},\label{eq:tens}
\end{eqnarray}
where $C=-2+\ln2+\gamma\approx-0.729637$ and $\gamma$ is the 
Euler constant. Although Eqs.~(\ref{eq:scal}-\ref{eq:tens}) are originally 
derived assuming $\epsilon$ and $\eta$ are constants~\cite{Stewart:1993bc}, 
evolution of these variables are practically very small at observable scales $k$, 
which validates the use of Eqs.~(\ref{eq:scal}-\ref{eq:tens})\footnote{
Several authors \cite{Hamann:2008pb,Adshead:2008vn}, 
have discussed effects of differences in the power 
spectra calculated from exact solution of the wave equation, approximation with
slow-roll parameters up to the 2nd order (i.e. Eqs.~(\ref{eq:scal}-\ref{eq:tens})). 
They conclude that differences among the exact and these approximated power 
spectra are not significant with a certain prior on the e-folding number $N$.
}.

The standard parameters of the primordial power spectra are given by Taylor 
expanding the logarithm of Eqs.~(\ref{eq:scal}-\ref{eq:tens}), 
\begin{eqnarray}
A_s&\equiv&P_\zeta(k_*), \label{eq:A_s}\\
n_s&\equiv&\left.\frac{d\ln\mathcal P_\zeta}{d\ln k}\right|_{k=k_*}+1
=1+2\eta_*-4\epsilon_*-2(1+\mathcal C)\epsilon_*^2-\frac{3-5\mathcal C}{2}
\epsilon_*\eta_*+\frac{3-\mathcal C}{2}\xi_*, \label{eq:n_s}\\
r&\equiv&\left.\frac{P_h}{P_\zeta}\right|_{k=k_*}=16\epsilon_*
(1+2 C(\epsilon_*-\eta_*)), \label{eq:r}\\
\alpha_s&\equiv&\left.\frac{d^2\ln\mathcal P_\zeta}{d\ln k^2}\right|_{k=k_*}
=-2\xi_*-8\epsilon_*^2+10\epsilon_*\eta_*, \label{eq:alpha_s}\\
n_t&\equiv&\left.\frac{d\ln\mathcal P_h}{d\ln k}\right|_{k=k_*}
=-2\epsilon_*-(3+\mathcal C)\epsilon_*^2+(1+\mathcal C)
\epsilon_*\eta_*, \label{eq:n_t}
\end{eqnarray}
where $\mathcal C= 4(\ln2+\gamma)-5$. With these standard parameters, we can
approximate the power spectra by
\begin{eqnarray}
\mathcal P_\zeta(k)&=&A_s\exp\left[(n_s-1)\ln\frac{k}{k_*}
+\frac{1}{2}\alpha_s\left(\ln\frac{k}{k_*}\right)^2\right], \label{eq:approx_s}\\
\mathcal P_h(k)&=&rA_s\exp\left[n_t\ln\frac{k}{k_*}\right]. \label{eq:approx_t}
\end{eqnarray}

Along with the power spectra in Eqs.~(\ref{eq:scal}-\ref{eq:tens}), the duration of the
inflation is also important. We impose a prior on the e-folding number, $N>25$. 
This prior roughly means that the energy scale of inflation should be higher
than TeV~\cite{Peiris:2006sj}\footnote{
The role of a prior on the e-folding number is 
detailedly discussed in \cite{Peiris:2008be}.
}.  Imposition of this prior ensures that inflaton 
does not affect physics below electroweak scales. 
However, this does not necessarily mean that a given model of single-field 
slow-roll inflation with certain ($\epsilon_*$, $\eta_*$, $\xi_*$) is excluded if it 
predicts $N<25$, for slow-roll parameters at higher orders can 
maintain slow-rolling of the inflaton further.
Thus we cannot strictly restrict models of single-field slow-roll inflation 
with a prior $N>25$. 
In Section~\ref{sec:analysis}, as default, we adopt a prior $N>25$ in investigating
constraints on the primordial power spectra with inflationary priors. However, 
we also investigate constraints without the prior $N>25$, which would be
informative in discussing what are plausible models supported from data.

\section{Bayesian model selection} 
\label{sec:selection}
As we have mentioned in Introduction, we adopt Bayesian model selection
in order to compare different models of single-field slow-roll inflation. Before
presenting details of our analysis, let us briefly review some essences of 
Bayesian model selection. We also refer to \cite{Trotta:2008qt,Mukherjee:2008zzb,
Liddle:2009xe} for more detailed reviews.

Given data, a joint posterior distribution for a model $M$ 
and its model parameter $\Theta$, $P(\Theta,M|\mathrm{data})$, is given by 
the hierarchical Bayes theorem, 
\begin{eqnarray}
P(\Theta,M|\mathrm{data})=\frac{P(\mathrm{data}|\Theta,M)P(\Theta|M)
P(M)}{P(\mathrm{data})}, \label{eq:jointpost}
\end{eqnarray}
where $P(\mathrm{data}|\Theta,M)=\mathcal L (\Theta)$ is the likelihood function,
$P(\Theta|M)$ is the prior distribution for $\Theta$ specified by $M$, 
and $P(M)$ is the prior distribution for $M$. 
The remaining $P(\mathrm{data})$ is an irrelevant normalization constant.
Then the posterior distribution 
for $M$, $P(M|\mathrm{data})$ is given by marginalizing 
$P(\Theta,M|\mathrm{data})$ over the model parameters $\Theta$,  
\begin{eqnarray}
P(M|\mathrm{data})&=&\int d\Theta
P(\Theta,M|\mathrm{data}) \notag\\ 
&=&\frac{P(M)}{P(\mathrm{data})}
\int d\Theta P(\mathrm{data}|\Theta,M)P(\Theta|M). \label{eq:modelpost}
\end{eqnarray}
The final integral in Eq.~(\ref{eq:modelpost}) is called Bayes evidence $E(M)$, 
\begin{eqnarray}
E(M)\equiv\int d\Theta P(\mathrm{data}|\Theta,M)P(\Theta|M), 
\label{eq:evidence}
\end{eqnarray}
which measures marginalized likelihood of the model $M$. Thus the relative likelihood of
different models $M_i$ and $M_j$ is assessed by the ratio of their Bayes evidences, 
\begin{eqnarray}
B_{ij}\equiv\ln\frac{E(M_i)}{E(M_j)} \label{eq:bayesfactor}
\end{eqnarray}
which is called Bayes factor. 
Bayesian model selection is implemented by estimating 
Bayes factor which measures relative likelihood between two different models.
Jeffreys' scale is often adopted to connect numbers and semantics, which states: 
for $|B_{ij}|<1$ the evidence is not significant; $1<|B_{ij}|<2.5$ significant; 
$2.5<|B_{ij}|<5$ strong; and $5<|B_{ij}|$ decisive\footnote{
The choice of numbers and semantics is more or less 
arbitrary and differs literature by literature. For instance, 
a more conservative statement is suggested in \cite{Trotta:2008qt}.}.

In usual parameter estimation, a posterior distribution of parameters
for a model $M$, $P(\Theta|\mathrm{data}, M)$ is investigated. However, 
our purpose in this paper of comparing different models of inflation is not achieved
via $P(\Theta|\mathrm{data}, M)$, where a model $M$ is by definition assumed
to be true. This is another way of representing our statement in Introduction, 
that different models cannot be compared with
a fixed parameter space. Instead, comparison of different models 
is achieved by the Bayes factors $B_{ij}$. This is why we adopt the Bayesian
model selection in our analysis. 
By assigning a prior $P(\Theta|M)$ to reflect prediction of a class of single-field inflation models $M$, 
the likelihood of $M$ is measured with the Bayes evidence $E(M)$ and two different classes can be compared. 

On the other hand, we can also explore an optimal constraint on parameters $\Theta$
by adopting Bayesian model selection. 
By comparing different models, which have separate priors on $\Theta$, 
an optimal constraints can be obtained from a posterior distribution  
$P(\Theta|\mathrm{data},M)$ of the parameters for a model 
which has higher Bayes evidence than other models.

One remark should be mentioned. The Bayes evidence explicitly depends on a prior
probability $P(\Theta|M)$ due to its normalization. For example let us consider 
that some two different top-hat priors are imposed on a same parameter. Even 
though the parameter region of high-likelihood is sufficiently covered by both of the 
priors, the resultant Bayes evidences differs, that are inversely proportional to the 
ranges of the top-hat priors. This is very contrastive to the usual parameter estimation,
where the posterior probabilities from these two different priors do not differ much.

\section{Analysis with observational data}
\label{sec:analysis}
Here we investigate constraints on the primordial
power spectra and make comparison of single-field slow-roll inflation models.
First of all, since there are quite a large number of theoretical models of single-field slow-roll
inflation, it is difficult to analyze them individually.
Instead we consider several different models
which have separate priors on the primordial power spectra.
Each of these models can be regarded as representing 
some class of single-field slow-roll inflation models.
We adopt two parameterizations of the primoridial 
power spectra. One is the standard parameterization of Eqs.
(\ref{eq:approx_s}-\ref{eq:approx_t}), that has $A_s$, $n_s$, $r$ and 
$\alpha_s$. As we are interested in single-field slow-roll inflation models, 
we impose the standard inflation consistency relation, $n_t=-r/8$, 
when we adopt the standard parameterization. The other is the HSR 
parameterization that has $A_s$, $\epsilon_*$, $\eta_*$ and $\xi_*$. 
By adopting these two parametrizations, the models we compare are listed in 
Table~\ref{tbl:models}. The reference model is $M_\mathrm{HZ}$ that has the 
HZ primordial power spectrum. $M_{n_s}$, $M_{n_sr}$ and $M_{n_s\alpha_s}$ have 
different top-hat priors on the standard parameters $n_s$, $r$ and $\alpha_s$, 
that are described in the last column of Table~\ref{tbl:models}. On the other hand, 
$M_{\epsilon}$, $M_{\eta}$, $M_{\eta\xi}$ have different top-hat priors on the HSR 
parameters  $\epsilon_*$ and $\eta_*$ that are also described in the last column of
Table~\ref{tbl:models}\footnote{
A similar division of models can also be found in \cite{Adshead:2008vn}.
}. In investigating models with the HSR parameters, we 
impose an additional prior $N>25$ as default. Basically, subscripts in names of 
models represent varied parameters in the model, and other parameters absent in
the subscripts are fixed to the default values. The only exception is $\epsilon_*$
which is varied in a small range $[0,10^{-4}]$ in $M_{\eta}$\footnote{
The upper bound $10^{-4}$ for $\epsilon_*$ is chosen so that these models cover
inflation scenarios where detection of primordial B-mode would be difficult with
CMB observations in the near future.
}. This is because $\epsilon_*$ cannot be fixed to zero as the inflaton cannot roll
when $\epsilon_*=0$. Regarding priors on $A_s$, we adopt a common prior 
$2.5<\ln[10^{10}A_s]<3.5$ in all models. 
In our analysis a fiducial wave number is chosen to be 
$k_*=0.01$~Mpc$^{-1}$\footnote{This is very near
a fiducial wave number $k_*=0.017$~Mpc$^{-1}$, 
which are suggested to be optimal for constraining 
the primordial power spectra from current data~\cite{Cortes:2007ak}.
Choice of fiducial wave number is also discussed in \cite{Peiris:2006sj}.}.

We assume a flat $\Lambda$CDM model as background cosmology. In addition to 
parameters representing the shape of the primordial power spectra 
(i.e. ($A_s$, $n_s$, $r$, $\alpha_s$) or ($A_s$, $\epsilon_*$, $\eta_*$, $\xi_*$)), following 
parameters are included in the analysis as primary cosmological parameters,  
\begin{eqnarray}
(\omega_b, \omega_c, \theta_s, \tau, A_{SZ}), 
\end{eqnarray}
where $\omega_b$ and $\omega_c$ are the density parameters of baryon and CDM, 
respectively, $\theta_s$ is the angular scale of the acoustic horizon~\cite{Kosowsky:2002zt}, 
$\tau$ is the optical depth of reionization, and $A_{SZ}$ 
is the amplitude of template Sunyaev-Zel'dovich power spectrum 
$C^{SZ}_\ell$~\cite{Komatsu:2002wc}. The priors on these primary cosmological 
parameters are listed in Table~\ref{tbl:priors}.

\begin{table}
  \begin{center}
  \begin{tabular}{l|c}
    \hline
    \hline
    models & description \\
    \hline
    $M_\mathrm{HZ}$ & HZ spectra \\
    \hline
    $M_{n_s}$ & $0.8<n_s<1.2$, $r=\alpha_s=0$\\ 
    $M_{n_sr}$ & $0.8<n_s< 1.2$, $0<r<0.5$, $\alpha_s=0$ \\
    $M_{n_s\alpha_s}$ & $0.8<n_s< 1.2$, $r=0$, $-0.1<\alpha_s<0.1$ \\
    \hline
    $M_{\epsilon}$ & $0<\epsilon_*< 0.1$, $\eta_*=\xi_*=0$ \\ 
    $M_{\eta}$ & $0<\epsilon_*< 10^{-4}$, $-0.1<\eta_*<0.1$, $\xi_*=0$ \\
    $M_{\epsilon\eta}$ & $0<\epsilon_*< 0.1$, $-0.1<\eta_*<0.1$, $\xi_*=0$ \\
    \hline
    \hline
  \end{tabular}
  \caption
  [Models adopted in the analysis]
  {Models adopted in the analysis. $M_\mathrm{HZ}$ is the reference model
  that has HZ primordial power spectrum. Other models cover a model space for
  single-field slow-roll inflation. $M_{n_s}$, $M_{n_sr}$, $M_{n_s\alpha_s}$ have
  top-hat priors on the standard parameters of the primordial power spectra,  
  and $M_\epsilon$, $M_\eta$, $M_{\epsilon\eta}$ have ones on the HSR 
  parameters. Prior ranges are described in the last column. Regarding priors on $A_s$, we adopt 
  a common prior $2.5<\ln[10^{10}A_s]<3.5$.
  }
  \label{tbl:models}
  \end{center}
\end{table}

We adopt CMB data from WMAP5 \cite{Dunkley:2008ie,Nolta:2008ih,
Hinshaw:2008kr}, as well as the observations at small angular scales including 
ACBAR \cite{Reichardt:2008ay}, CBI \cite{Sievers:2005gj}, 
BOOMERANG~\cite{Jones:2005yb,Piacentini:2005yq,Montroy:2005yx} 
and QUAD \cite{Friedman:2009dt}. In addition, we adopt observational data of 
geometrical distances of the late Universe, including the Union data set of 
SN \cite{Kowalski:2008ez}, the measurement of BAO 
scales in galaxy power spectra \cite{Percival:2009xn}\footnote{
We also performed the same analysis by adopting the data of halo power spectra from
the catalogue of the SDSS Luminous Red Galaxies \cite{Reid:2009xm} instead of the
BAO data. However, as long as combined with data of CMB, SN and the $H_0$ 
measurement, the results are very similar to those from the BAO data, which is also
consistently observed in Ref. \cite{Finelli:2009bs}. 
We suppose that this is due to the marginalization over scale-dependent galaxy biases. 
Although the matter power spectrum itself is a promising probe for the primordial power
spectrum at small scales (See also \cite{Peiris:2009wp} for a recent result), 
there are some difficulties in modeling nonlinear evolution 
of matter perturbations, biases of tracers, or effects of baryonic physics.
For the time being, we do not adopt data of the matter power spectrum.
} and the SH$_0$ES measurement of the Hubble constant 
$H_0=74.2\pm3.6$ \cite{Riess:2009pu}.
A default data set that we conduct our analysis with is a combination of all the 
data above, which we denote as `ALL'. On the other hand, we also make use 
of several other data sets, in order to assure ourselves that the results are not 
dragged by some extreme data. Adopted are data sets of WMAP5 alone and 
the combinations of WMAP5 with either the other CMB data at small angular 
scales, BAO, SN, or the $H_0$ measurement, which we denote as +CMB, 
+BAO, +SN, +$H_0$, respectively.

\begin{table}
  \begin{center}
  \begin{tabular}{l|c}
    \hline
    \hline
    parameters & prior ranges \\
    \hline
    $\omega_b$ & $[0.02,0.025]$ \\
    $\omega_c$ & $[0.08,0.14]$ \\
    $\theta_s$ & $[1.02,1.06]$ \\
    $\tau$ & $[0.01,0.2]$ \\
    $A_{SZ}$ & $[0,4]$ \\
    \hline
    \hline
  \end{tabular}
  \caption
  [Top-hat priors on other cosmological parameters]
  {Top-hat priors on cosmological parameters, 
  other than parameters of the primordial power spectra.}
  \label{tbl:priors}
  \end{center}
\end{table}

Computation of Bayes evidence is implemented with
{\tt MultiNest}~\cite{Feroz:2008xx}, which is integrated in the Markov chain Monte Carlo 
(MCMC) sampling code {\tt CosmoMC}~\cite{Lewis:2002ah} but uses the nested 
sampling~\cite{Skilling:2004} in stead of the MCMC sampling (See also 
Ref.~\cite{Mukherjee:2005wg}). Given a model $M$, {\tt MultiNest} provides 
chains of samples from the posterior distribution $P(\Theta|\mathrm{data},M)$ and
Bayes evidence $E(M)$. Convergence is diagnosed
by applying the Gelman-Rubin test with two independent chains. 
Typically $R-1<0.01$ is achieved.

\subsection{Standard parameterization}
\label{subsec:standard}
\begin{figure}
  \begin{center}
    \begin{tabular}{ccc}
      \hspace{-10mm}\scalebox{0.8}{\includegraphics{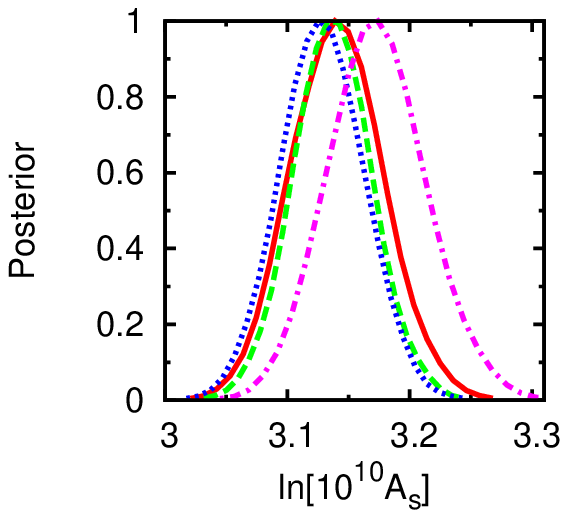}} \\
      \hspace{-10mm}\scalebox{0.8}{\includegraphics{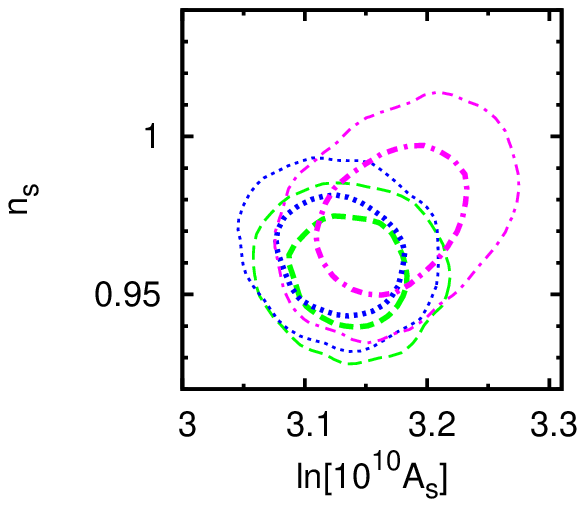}} &
      \hspace{-10mm}\scalebox{0.8}{\includegraphics{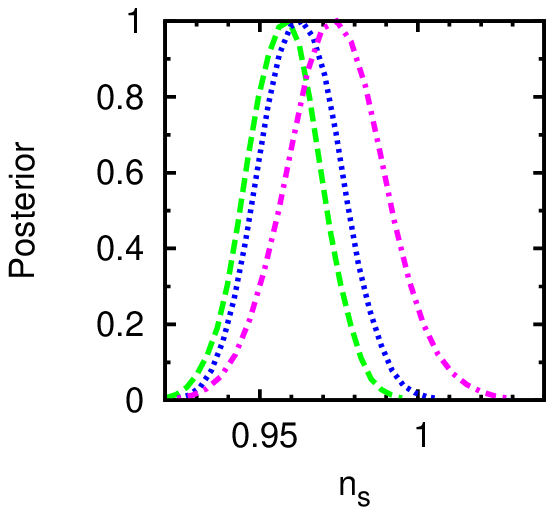}} \\
      \hspace{-10mm}\scalebox{0.8}{\includegraphics{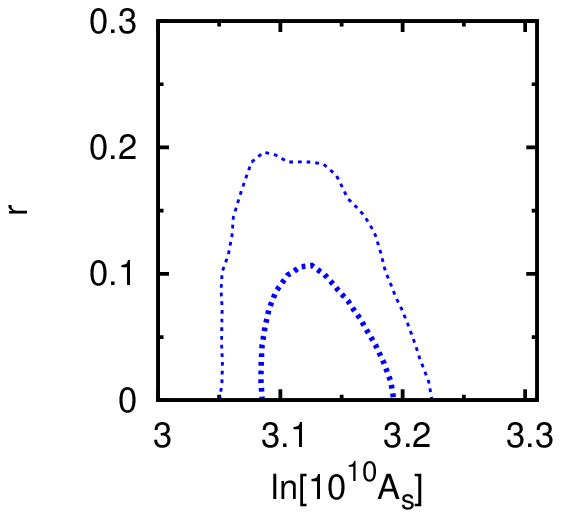}} &
      \hspace{-10mm}\scalebox{0.8}{\includegraphics{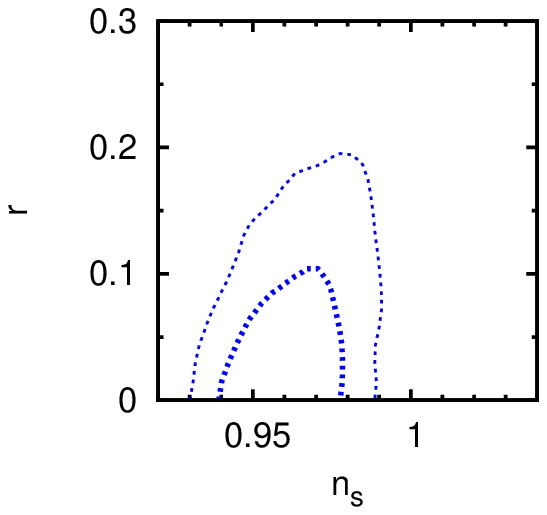}} &
      \hspace{-10mm}\scalebox{0.8}{\includegraphics{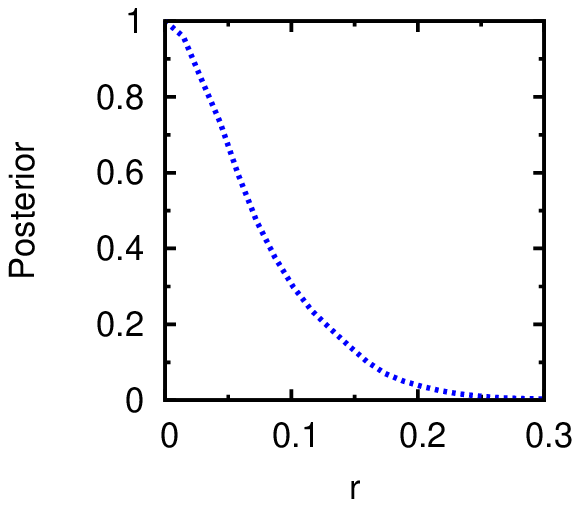}} \\
      \hspace{-10mm}\scalebox{0.8}{\includegraphics{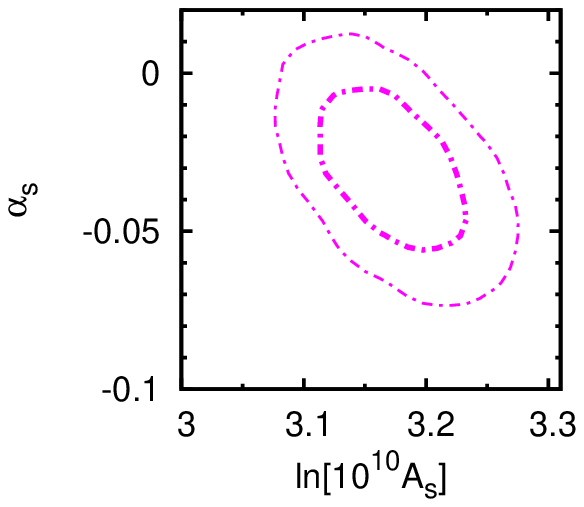}} &
      \hspace{-10mm}\scalebox{0.8}{\includegraphics{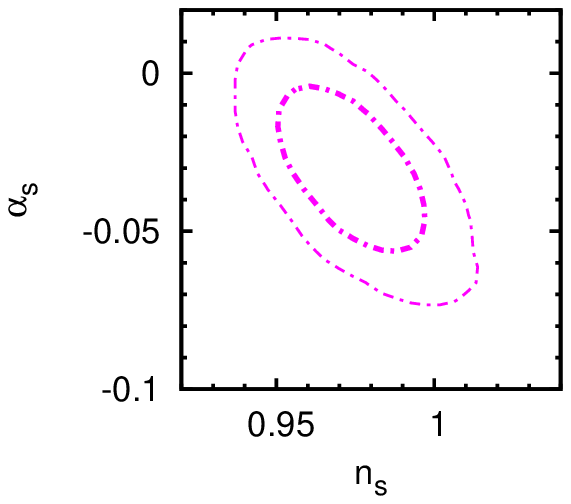}} &
      \hspace{-10mm}\scalebox{0.8}{\includegraphics{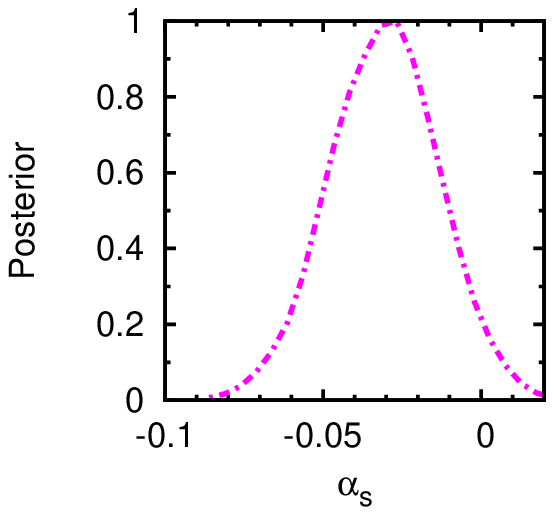}} \\
    \end{tabular}
  \end{center}
  \caption{1d and 2d marginalized posterior distributions of the standard parameters, 
  $\log[10^{10}A_s]$, $n_s$, $r$ and $\alpha_s$ from the ALL data set.
  Models shown are $M_{n_s}$ (dashed green line), $M_{n_sr}$ (dotted blue line),  
  $M_{n_s\alpha_s}$ (dot-dashed magenta line) and the reference $M_\mathrm{HZ}$ (solid red line).}
  \label{fig:standard}
\end{figure}

In Table \ref{tbl:standard} we list Bayes factors for 
models with standard parameterization of the primordial power spectra.
$M_{n_s}$, $M_{n_sr}$ and $M_{n_s\alpha_s}$ against a
reference model $M_\mathrm{HZ}$. 
What we may first see in the table is that
almost all data sets give positive Bayes factors.
Although there are a few exceptions given by the $+H_0$ data set,
they are by no means significant.
This result shows us that current 
data negatively support a model with the HZ spectrum.
In particular, the strongest negative support is 
brought from the ALL data set.
Although Bayes factors depend on models we compare with $M_\mathrm{HZ}$,  
all models we adopt, $M_{n_s}$, $M_{n_sr}$ and $M_{n_s\alpha_s}$, 
give Bayes factors larger than 2.5. On Jeffreys' scale, these results 
correspond to strong negative evidences for $M_\mathrm{HZ}$.
Such negative preference for the HZ power spectra was
reported by many previous studies~\cite{Trotta:2005ar,Kunz:2006mc,
Parkinson:2006ku,Bridges:2006zm,Liddle:2006tc,Gordon:2007xm}
and our result is consistent with the other recent result from WMAP5 \cite{Bridges:2008ta}.
However, significance in our result is considerably large as very recent 
observations are adopted. 
In Figure~\ref{fig:standard}, we plotted the 1d and 2d marginalized 
posterior distributions of the parameters $n_s$, $r$, and $\alpha_s$
from the ALL data set. 
Clearly, the negative support for $M_\mathrm{HZ}$ originates from a
poor fit of the HZ spectrum to the data, as the posterior distributions 
of $n_s$ for models $M_{n_s}$, $M_{n_sr}$ and $M_{n_s\alpha_s}$
consistently show preference for $n_s\ne1$.

The Bayes evidences in Table~\ref{tbl:standard} also show that the current data 
can be well-modeled by scalar perturbation with power-law power spectrum 
without the tensor mode (i.e. $M_{n_s}$).
Let us regard $M_{n_s}$ as a base model and ALL as a base data set.
When we include the tensor perturbation by varying $r>0$, 
the Bayes factors decrease from $4.3$ for $M_{n_s}$ to $2.6$ for $M_{n_sr}$. 
This suggests that current data negatively support presence of the
tensor perturbation. In fact, the posterior distribution for $r$ 
peaks at $r=0$, as seen in Figure~\ref{fig:standard}.
On the other hand, when we include the running of the scalar spectral index, the
Bayes factor is almost unchanged from 4.3 for $M_{n_s}$ to 4.4 for $M_{n_s\alpha_s}$.
This shows that even though the posterior distribution of $\alpha_s$ 
in Figure~\ref{fig:standard} peaks away from $\alpha_s=0$, 
the likelihood does not improve much from $M_{n_s}$ in  
most region of $\alpha_s\ne0$ for the $M_{n_s\alpha_s}$ model.
We conclude that presence of $\alpha_s$ is not disfavored 
by the data nor required in modeling the data.

From Table~\ref{tbl:standard}, the preference for $M_{n_s}$ over $M_{n_sr}$ 
is commonly observed with all the data sets.
Also $M_{n_s}$ and $M_{n_s\alpha_s}$ are supported almost 
equally since the Bayesian evidences for these models do not differ by 1 at most.
Totally, as $M_{n_s}$ is nested by $M_{n_sr}$ or $M_{n_s\alpha_s}$, 
we can safely conclude that $M_{n_s}$ best describes current data
as far as we compare models with standard parameterization listed in Table~\ref{tbl:standard}.

The constraints on cosmological parameters for these models are summarized 
in Table~\ref{tbl:standard_constr}. As seen in the table, the current data can  
optimally be modeled by $M_{n_s}$. 
Therefore, we propose the constraint for $M_{n_s}$ as 
optimal one\footnote{
We do not derive an optimal constraint by averaging over models, 
as our division of models, $M_{n_s}$, $M_{n_sr}$ and $M_{n_s\alpha_s}$,
is rather artificial, and not strongly based on theoretical motivations.
}. An optimal constraint on the primordial power spectra 
from the ALL data set is given by
\begin{eqnarray}
\ln[10^{10}A_s]&=&3.137^{+0.028}_{-0.032},
\label{eq:opt_A_s} \\
n_s&=&0.957^{+0.010}_{-0.011}, 
\label{eq:opt_n_s}
\end{eqnarray}
where errors are given at 68\% C.L. 
Regarding the tensor perturbation and the running of the scalar index,
we find no evidence at present.

As a final remark of this section, let us comment on the 
choice of prior distributions and its relations to the resultant Bayes evidences.
As we have mentioned in Section~\ref{sec:selection}, 
Bayes evidences depend on prior distributions.
While a prior $0.8<n_s<1.2$ is often adopted in other studies 
\cite{Parkinson:2006ku,Gordon:2007xm}, 
priors $r<0.5$ and $|\alpha_s|<0.1$ may look moderately tight. 
More relaxed priors on $r$ and $\alpha_s$ may slightly decrease 
the Bayes evidences for $M_{n_sr}$ and $M_{n_s\alpha_s}$,
which however hardly affects our conclusion. 

\begin{table}
  \begin{center}
  \begin{tabular}{lrrr}
    \hline
    \hline
    models & $M_{n_s}$ & $M_{n_sr}$ & $M_{n_s\alpha_s}$ \\
    \hline
    WMAP5 & \ \ $+0.8\pm0.2$ & \ \ $-0.1\pm0.2$& \ \ $+0.3\pm0.2$\\
    \hline
    \ \ +CMB & \ \ $+1.4\pm0.2$ & \ \ $+0.5\pm0.2$& \ \ $+1.2\pm0.2$\\
    \hline
    \ \ +BAO & \ \ $+1.7\pm0.2$ & \ \ $+0.4\pm0.2$& \ \ $+1.5\pm0.2$\\
    \hline
    \ \ +SN & \ \ $+1.6\pm0.2$ & \ \ $+0.3\pm0.2$& \ \ $+1.9\pm0.2$\\
    \hline
    \ \ +$H_0$ & \ \ $+0.5\pm0.2$ & \ \ $-0.2\pm0.2$& \ \ $-0.4\pm0.2$\\
    \hline
    ALL & \ \ $+4.3\pm0.2$ & \ \ $+2.6\pm0.2$& \ \ $+4.4\pm0.2$\\
    \hline
    \hline
  \end{tabular}
  \caption{Bayes factors
  for models with the standard parameters of the primordial power spectra 
  against the reference model $M_\mathrm{HZ}$. 
  }
  \label{tbl:standard}
  \end{center}
\end{table}

\begin{table}
  \begin{center}
  \begin{tabular}{lrrr}
    \hline
    \hline
    parameters & $M_{n_s}$ & $M_{n_sr}$ & $M_{n_s\alpha_s}$  \\
    \hline
    $\omega_b\times10^2$ & $2.263^{+0.055}_{-0.041}$ & $2.278^{+0.051}_{-0.048}$ & $2.227^{+0.044}_{-0.059}$ \\
    $\omega_c$ & $0.1133^{+0.0028}_{-0.0032}$ & $0.1128^{+0.0027}_{-0.0033}$ & $0.1155^{+0.0032}_{-0.0036}$ \\
    $\theta_s\times10^2$ & $104.09^{+0.19}_{-0.24}$ & $104.10^{+0.22}_{-0.19}$ & $104.12^{+0.21}_{-0.21}$ \\
    $\tau$ & $0.085^{+0.013}_{-0.017}$ & $0.084^{+0.015}_{-0.015}$ & $0.095^{+0.015}_{-0.020}$ \\
    $\ln[10^{10}A_s]$ & $3.137^{+0.028}_{-0.032}$ & $3.128^{+0.030}_{-0.035}$ & $3.173^{+0.034}_{-0.042}$ \\
    $n_s$ & $0.957^{+0.010}_{-0.011}$ & $0.962^{+0.012}_{-0.011}$ & $0.974^{+0.013}_{-0.015}$ \\
    $r$ & --- & $(0,0.15)$ & --- \\
    $\alpha_s$ & --- & --- & $-0.030^{+0.18}_{-0.14}$ \\
    $n_t$ & --- & $(-0.02,0)$ & --- \\
    \hline
    \hline
  \end{tabular}
  \caption{Constraints for models with the standard parameters of the primordial power spectra, 
  $M_{n_s}$, $M_{n_sr}$ and $M_{n_s\alpha_s}$, from the ALL data set. 
  For bounded parameters, shown are mean values and 68\% credible intervals.
  For unbounded parameters, 95\% credible intervals are only shown with parentheses.
  }
  \label{tbl:standard_constr}
  \end{center}
\end{table}

\subsection{HSR parameterization}
\label{subsec:slow-roll}
In Tables~\ref{tbl:HSR} we present the Bayes factors for models with the HSR 
parameters, $M_\epsilon$, $M_\eta$ and $M_{\epsilon\eta}$, against $M_\mathrm{HZ}$.
From the table, first we reconfirm that the $M_\mathrm{HZ}$ model is negatively supported from the data.
With the ALL data set, the Bayes factors are $5.8$, $4.7$ and $4.2$ for 
$M_\epsilon$, $M_\eta$ and $M_{\epsilon\eta}$, respectively.
On Jeffreys' scale, these correspond to negative 
support for the model with the HZ spectrum at decisive or at least strong level. 
This shows that models of primordial power spectra with inflationary priors 
can quite well explain the data, while the data are very poorly fitted by the HZ spectrum.

\begin{figure}
  \begin{center}
    \begin{tabular}{ccc}
      \hspace{-10mm}\scalebox{0.8}{\includegraphics{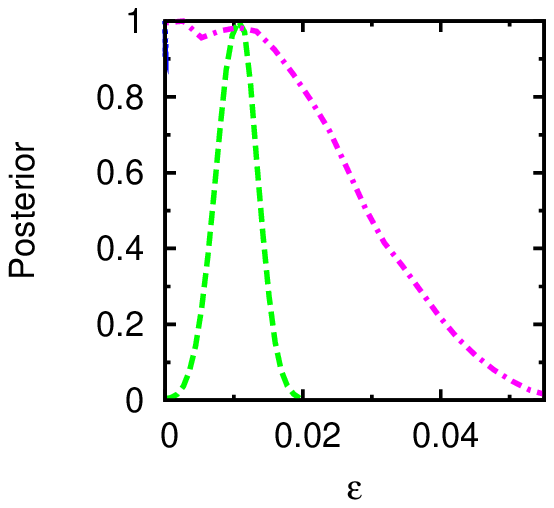}} & \\
      \hspace{-10mm}\scalebox{0.8}{\includegraphics{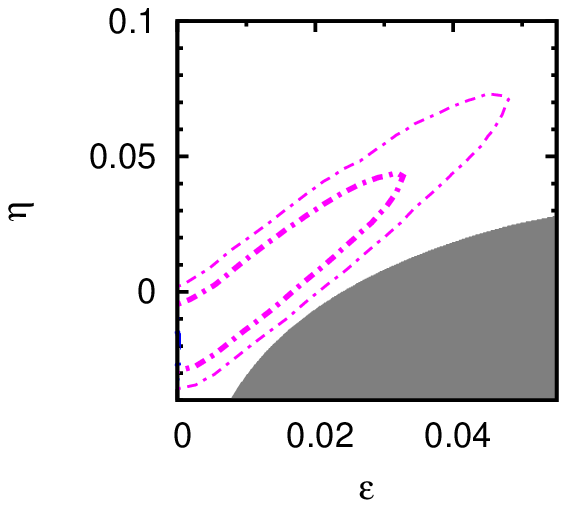}} &
      \hspace{-10mm}\scalebox{0.8}{\includegraphics{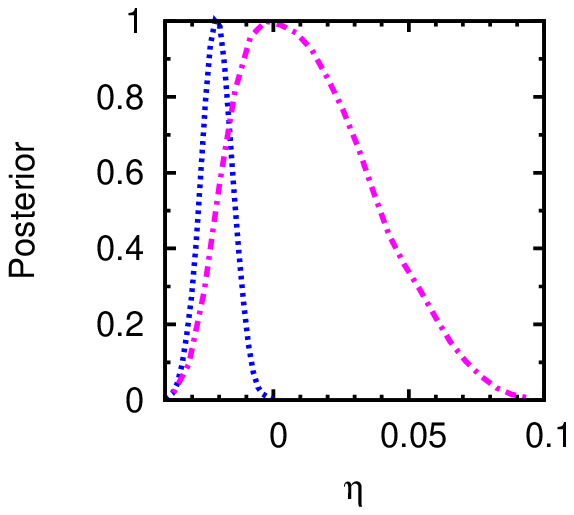}} \\
      \hspace{-10mm}\scalebox{0.8}{\includegraphics{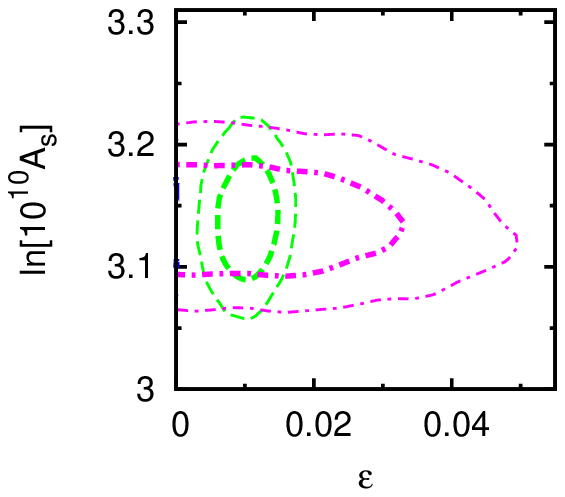}} &
      \hspace{-10mm}\scalebox{0.8}{\includegraphics{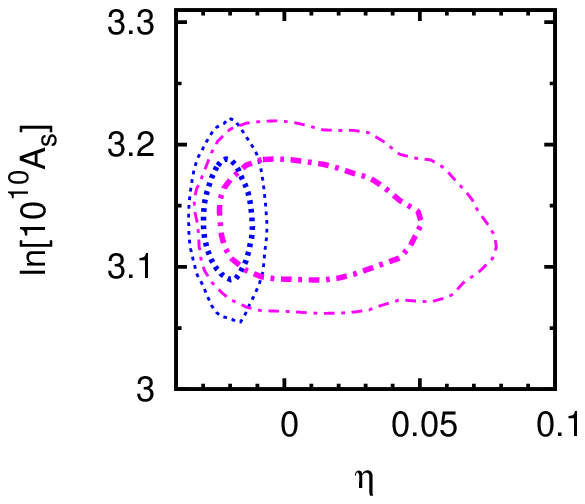}} &
      \hspace{-10mm}\scalebox{0.8}{\includegraphics{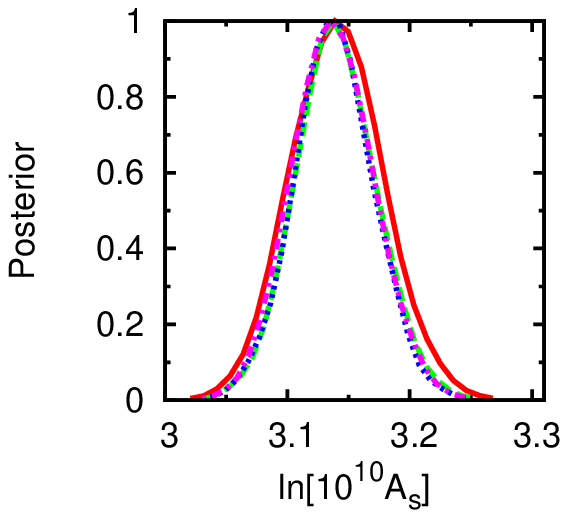}} \\
    \end{tabular}
  \end{center}
  \caption{1d and 2d marginalized posterior distributions for the HSR parameters, $\epsilon_*$ and $\eta_*$,
  and the amplitude of the curvature power spectrum $\ln[10^{10}A_s]$. 
  Models shown are $M_\epsilon$ (dashed green line), $M_\eta$ (dotted blue line), 
  $M_{\epsilon\eta}$ (dot-dashed magenta line) and the reference $M_\mathrm{HZ}$ (solid red line). 
  The gray shaded region is excluded by a prior $N>25$ when $\xi_*=0$.}
  \label{fig:HSR}
\end{figure}

\begin{figure}
  \begin{center}
    \begin{tabular}{cccc}
      \hspace{-10mm}\scalebox{0.6}{\includegraphics{inf_all_8.eps}} \\
      \hspace{-10mm}\scalebox{0.6}{\includegraphics{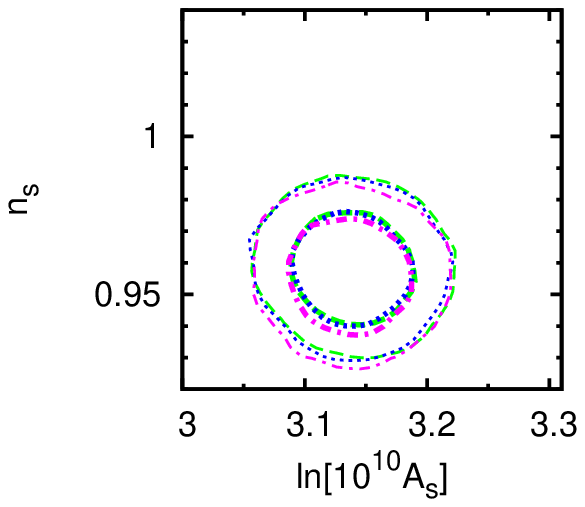}} &
      \hspace{-10mm}\scalebox{0.6}{\includegraphics{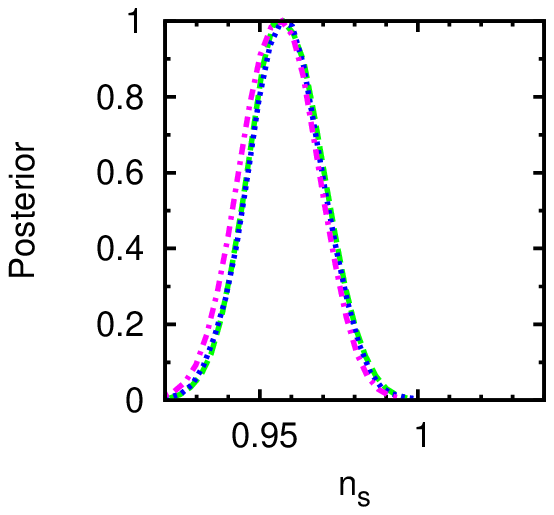}} \\
      \hspace{-10mm}\scalebox{0.6}{\includegraphics{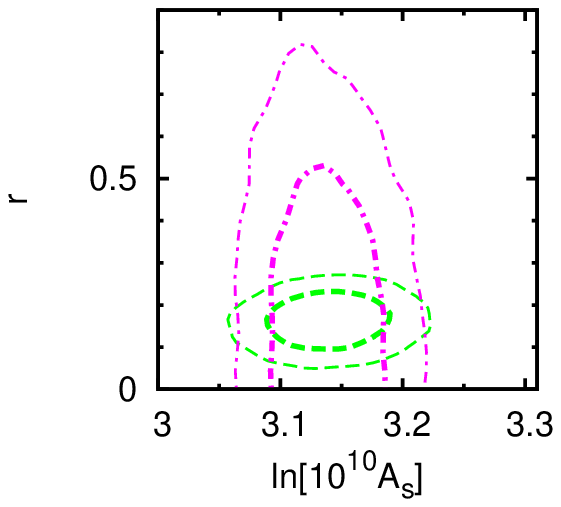}} &
      \hspace{-10mm}\scalebox{0.6}{\includegraphics{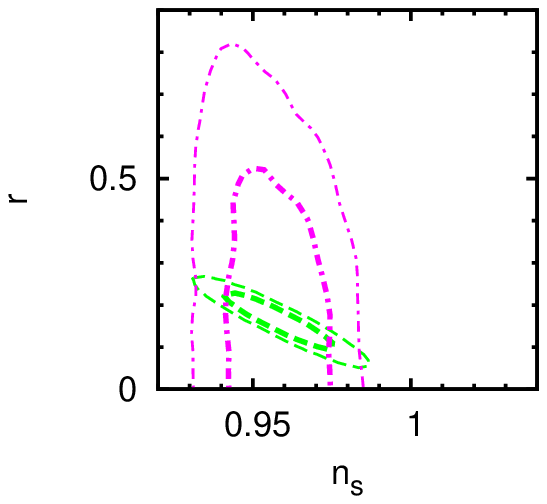}} &
      \hspace{-10mm}\scalebox{0.6}{\includegraphics{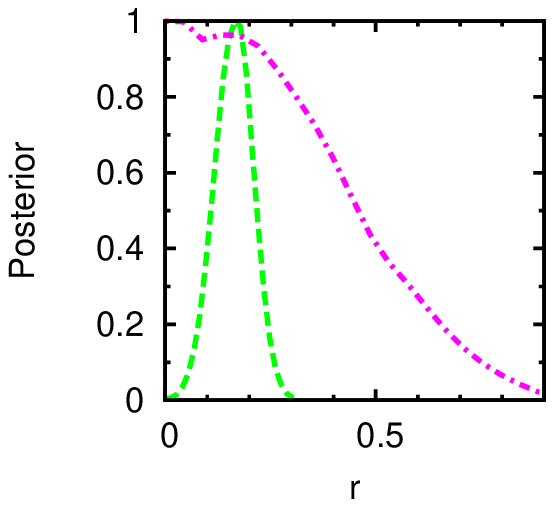}} \\
      \hspace{-10mm}\scalebox{0.6}{\includegraphics{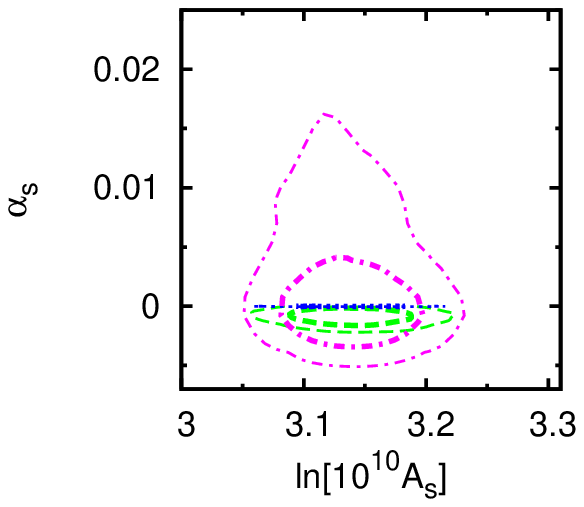}} &
      \hspace{-10mm}\scalebox{0.6}{\includegraphics{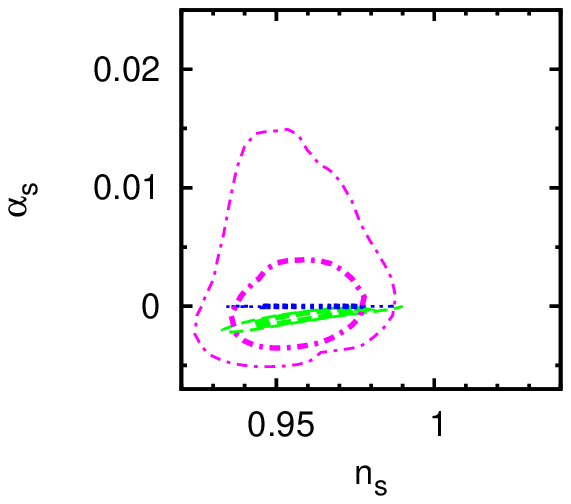}} &
      \hspace{-10mm}\scalebox{0.6}{\includegraphics{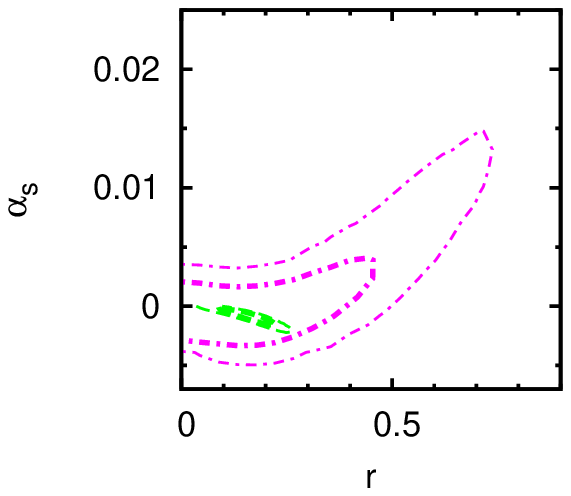}} &
      \hspace{-10mm}\scalebox{0.6}{\includegraphics{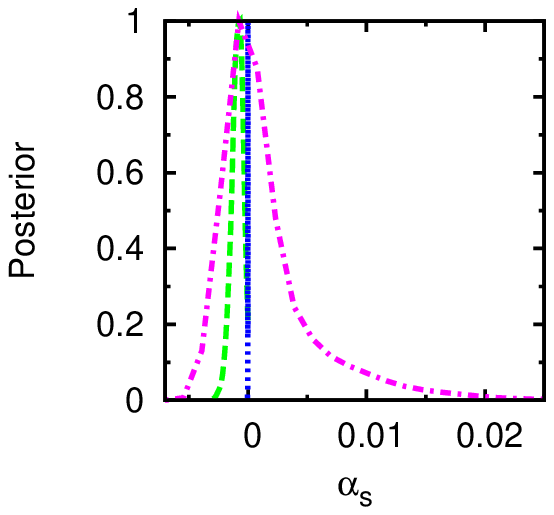}} \\
    \end{tabular}
  \end{center}
  \caption{Same figure as in Figure~\ref{fig:HSR} but the 
  posterior distributions for the 
  standard parameters, $\ln[10^{10}A_s]$, $n_s$, $r$ and $\alpha_s$, are shown.}
  \label{fig:HSR_derived}
\end{figure}

Let us take a close look at the results so as to find plausible models 
of single-field slow-roll inflation from current data.
In Figure~\ref{fig:HSR}, we presented the marginalized 1d and 2d posterior 
distributions of the HSR parameters $\epsilon_*$ and $\eta_*$. 
There we explicitly show the parameter region in $\epsilon_*$-$\eta_*$ plane 
that are removed by a prior $N>25$ for $\xi_*=0$. 
In Figure \ref{fig:HSR_derived}, we also 
present the marginalized posterior distributions of the amplitude 
of curvature power spectrum $A_s$ and  the {\it derived} standard parameters, 
$n_s$, $r$ and $\alpha_s$, given as functions of the HSR parameters 
(See Eqs. (\ref{eq:n_s}-\ref{eq:alpha_s})).
From Table~\ref{tbl:HSR} we consistently observe that the 
Bayes evidences for $M_\epsilon$ are larger than those for
$M_\eta$ and $M_{\epsilon\eta}$, which are nearly equal, 
for all the data sets we adopted. With the ALL data set, the 
Bayes factor of $M_\epsilon$ against $M_\eta$ and 
$M_{\epsilon\eta}$ are $1.1$ and $1.6$. Thus the current data 
show slight preference for $M_\epsilon$ over $M_{\eta}$ and 
$M_{\epsilon\eta}$, which would be significant on Jeffreys' scale.

However, the preference for $M_\epsilon$ looks somewhat dependent on 
several assumptions we adopted in the analysis and the origin is not clear. 
One possible origin would be the prior $N>25$, for 
it eliminates the region $\epsilon_*\gtrsim0.024$ if $\eta_*=0$, as can be seen
from the shaded region in Figure \ref{fig:HSR}. Thus in the models of $M_\epsilon$, 
the original top-hat prior $0<\epsilon_*<0.1$ is effectively substituted with a top-hat prior 
$0<\epsilon_*<0.024$ due to the prior $N>25$.
As can be seen in the marginal posterior distribution of $\epsilon_*$ for 
$M_\epsilon$ in Figure \ref{fig:HSR}, the effective prior region appears to 
give high likelihood, which results in boosting up the Bayesian evidence for $M_\epsilon$.

Let us examine the origin of preference for $M_\epsilon$ in more detail.
We have repeated Monte Carlo analyses using the ALL data set, 
modifying the default setup in several different ways as follows:
\begin{enumerate}
\item[(1)] A prior $N>25$ is removed.
\item[(2)] Contribution from the tensor perturbation is omitted in the power spectra of CMB anisotropies.
\item[(3)] The prior ranges for $\epsilon_*$ and $\eta_*$ are extended 
from $\epsilon_*,|\eta_*|<0.1$ to $\epsilon_*,|\eta_*|<0.2$.
\item[(4)] The higher order HSR parameter $\xi_*$ is 
included with a top-hat prior $-0.01<\xi_*<0.01$.
\end{enumerate}
The results are summarized in Table~\ref{tbl:extension}. 
From the modification (1), we first note that the Bayes fractor for 
$M_\epsilon$ significantly decreases from $5.8$ to $4.7$ by removal of a prior $N>25$.
The resultant Bayes evidences are very comparable between models $M_\epsilon$ and $M_\eta$, 
while more complicated $M_{\epsilon\eta}$ gives a smaller Bayes evidence.

On the other hand, the largest difference between 
a model with large $\epsilon_*$ (e.g. $M_\epsilon$) 
and one with small $\epsilon_*$ (e.g. $M_\eta$) 
would be whether primordial tensor perturbation is generated or not.
Therefore the preference for $M_\epsilon$ can also 
possibly be induced from observably large contribution 
of the tensor perturbation in CMB anisotropies. However, 
this possibility is not supported by the result from 
the modification (2). The Bayes factor for $M_\epsilon$ 
does not decrease by omission of the tensor contribution,
or even comes to increase slightly 
from $6.0$ to $6.3$. This is also as expected from the 
discussions in Section~\ref{subsec:standard},  where the 
model $M_{n_sr}$ with nonzero $r$ is less supported, 
compared with $M_{n_s}$ with vanishing $r$. From 
Figure~\ref{fig:HSR_derived} we also note that the 
running of scalar spectral index for these models 
is too small to be observed with current data 
(See also constraints on $\alpha_s$ for $M_{n_s\alpha_s}$ in Figure~\ref{fig:standard}). 
Therefore the preference for $M_{\epsilon}$ is not 
from the running of the scalar spectral index either.

With all above demonstrations, we conclude that 
the preference for $M_\epsilon$ originates from 
a theoretical prior $N>25$. Current data is not very 
constraining enough for us to distinguish models adopted in 
our analysis and future observations should be await.

We also examine the dependence of Bayes evidence
on top-hat priors on $\epsilon_*$ and $\eta_*$. 
The modification (3) in Table~\ref{tbl:extension} shows that 
doubling of prior ranges from $\epsilon_*, |\eta_*|<0.1$ to $\epsilon_*, |\eta_*|<0.2$
decreases the Bayes factors for $M_\eta$, $M_{\epsilon\eta}$ by $0.8$ and $1.4$, 
respectively. These decreases are almost as expected from the dependence of Bayes 
evidences on prior ranges, that are about $\ln 2\simeq0.7$ and $2\ln2\simeq1.4$ 
for $M_\eta$ and $M_{\epsilon\eta}$, respectively. 
Contrastively, the Bayes evidence for $M_\epsilon$ is unchanged by the modification (3)
because the priors $N>25$ dominates over the original top-hat prior on $\epsilon_*$.

So far we restricted ourselves to models with the HSR parameters up to 
$\epsilon$ and $\eta$. Let us examine possible preference for models 
with higher order HSR parameters. 
We have included the third lowest order HSR parameter $\xi_*$ 
with a top-hat prior $-0.01<\xi_*<0.01$ in the modification (4).
From Table~\ref{tbl:extension} we observe that 
the Bayes evidences almost equal or decrease from those with vanishing $\xi_*$.
Therefore current data show no preference for models with nonzero $\xi_*$, 
and probably models with further higher order HSR parameters, as well. 
Here we should remark that above conclusion for models with nonzero $\xi_*$ relies on a prior $N>25$. 
As explicitly shown in~\cite{Peiris:2006sj}, positive $\xi_*\simeq\mathcal O(0.01)$ 
may be favored when we omit the prior $N>25$. 
As we have discussed in Section~\ref{sec:HSR}, even for such large 
$\xi_*$, a sufficient e-folding number $N>25$ can possibly be achieved by 
higher order HSR parameters or other subsequent inflations. 
However, such complication may make the models less appealing.

The constraints on cosmological parameters for the models with 
the HSR parameters are presented in Table~\ref{tbl:HSR_constr}. 
From the table, we see that constraints on the amplitude and the spectral index of the 
primordial curvature power spectrum, $A_s$ and $n_s$, 
are almost independent of priors on the HSR parameters 
(See also Figure~\ref{fig:HSR_derived}). Moreover, these constraints 
are identical with those of Eqs. (\ref{eq:opt_A_s}-\ref{eq:opt_n_s}) from the $M_{n_s}$ models. 
Therefore we conclude that the optimal constraint we have proposed in Section~\ref{subsec:standard} 
is robust with little dependence on priors from different inflationary models.
However, constraints on the tensor perturbations and the running of the spectral index 
are largely dependent on inflationary priors and their presence is not suggested 
from current data. 

\begin{table}
  \begin{center}
  \begin{tabular}{lrrr}
    \hline
    \hline
    models & $M_{\epsilon}$ & $M_{\eta}$ & $M_{\epsilon\eta}$ \\
    \hline
    WMAP5 & \ \ $+2.0\pm0.2$ & \ \ $+0.9\pm0.2$& \ \ $+0.6\pm0.2$\\
    \hline
    \ \ +CMB & \ \ $+2.9\pm0.2$ & \ \ $+1.7\pm0.2$& \ \ $+1.1\pm0.2$\\
    \hline
    \ \ +BAO & \ \ $+3.2\pm0.2$ & \ \ $+1.9\pm0.2$& \ \ $+1.6\pm0.2$\\
    \hline
    \ \ +SN & \ \ $+3.0\pm0.2$ & \ \ $+1.9\pm0.2$& \ \ $+1.7\pm0.2$\\
    \hline
    \ \ +$H_0$ & \ \ $+1.8\pm0.2$ & \ \ $+0.6\pm0.2$& \ \ $+0.3\pm0.2$\\
    \hline
    ALL & \ \ $+5.8\pm0.2$ & \ \ $+4.7\pm0.2$& \ \ $+4.2\pm0.2$\\
    \hline
    \hline
  \end{tabular}
  \caption{Bayes factors for models with the HSR parameters
  against the reference model $M_\mathrm{HZ}$. 
  }
  \label{tbl:HSR}
  \end{center}
\end{table}

\begin{table}
  \begin{center}
  \begin{tabular}{lrrr}
    \hline
    \hline
    models & $M_{\epsilon}$ & $M_{\eta}$ & $M_{\epsilon\eta}$ \\
    \hline
    ALL & \ \ $+5.8\pm0.2$ & \ \ $+4.7\pm0.2$& \ \ $+4.2\pm0.2$\\
    \hline
    (1) & \ \ $+4.7\pm0.2$ & \ \ $+4.6\pm0.2$& \ \ $+3.7\pm0.2$\\
    \hline
    (2) & \ \ $+5.9\pm0.2$ & \ \ $+4.6\pm0.2$& \ \ $+4.2\pm0.2$\\
    \hline
    (3) & \ \ $+5.8\pm0.2$ & \ \ $+3.9\pm0.2$& \ \ $+2.8\pm0.2$\\
    \hline
    (4) & \ \ $+4.3\pm0.2$ & \ \ $+5.1\pm0.2$& \ \ $+3.6\pm0.2$\\
    \hline
    \hline
  \end{tabular}
  \caption{Bayes factors for models with the HSR prameters
  against the reference model $M_\mathrm{HZ}$.
  Several extensions from the default setting are investigated: 
  (1) removal of a prior on e-folding number;
  (2) omission of the tensor perturbation;
  (3) imposition of extended top-hat priors, $\epsilon_*,|\eta_*|<0.2$;
  (4) inclusion of the third lowest order HSR parameter $-0.01<\xi_*<0.01$.
  }
  \label{tbl:extension}
  \end{center}
\end{table}

\begin{table}
  \begin{center}
  \begin{tabular}{lrrr}
    \hline
    \hline
    parameters & $M_{\epsilon}$ & $M_{\eta}$ & $M_{\epsilon\eta}$  \\
    \hline
    $\omega_b\times10^2$ & $2.261^{+0.052}_{-0.048}$ & $2.263^{+0.048}_{-0.049}$ & $2.260^{+0.051}_{-0.050}$ \\
    $\omega_c$ & $0.1136^{+0.0031}_{-0.0031}$ & $0.1135^{+0.0029}_{-0.0032}$ & $0.1134^{+0.0029}_{-0.0032}$ \\
    $\theta_s\times10^2$ & $104.12^{+0.22}_{-0.19}$ & $104.13^{+0.21}_{-0.20}$ & $104.11^{+0.23}_{-0.19}$ \\
    $\tau$ & $0.086^{+0.013}_{-0.017}$ & $0.085^{+0.015}_{-0.014}$ & $0.084^{+0.012}_{-0.018}$ \\
    $\epsilon_*$ & $0.0104^{+0.0028}_{-0.0028}$ & --- & $(0,0.039)$ \\
    $\eta_*$ & --- & $-0.0208^{+0.0059}_{-0.0051}$ & $0.013^{+0.015}_{-0.030}$ \\
    $\ln[10^{10}A_s]$ & $3.137^{+0.028}_{-0.032}$ & $3.137^{+0.030}_{-0.030}$ & $3.137^{+0.027}_{-0.036}$ \\
    \hline
    $n_s$ & $0.958^{+0.011}_{-0.011}$ & $0.958^{+0.011}_{-0.010}$ & $0.956^{+0.011}_{-0.012}$ \\
    $r$ & $0.163^{+0.044}_{-0.043}$ & --- & $(0,0.64)$ \\
    $\alpha_s\times 10^3$ & $0.92^{+0.57}_{-0.32}$ & --- & $1.57^{+0.06}_{-3.20}$ \\
    $n_t\times10^2$ & $-2.10^{+0.58}_{-0.58}$ & --- & $(-8.1,0)$ \\
    \hline
    \hline
  \end{tabular}
  \caption{Constraints for models with the inflationary HSR parameters, 
  $M_{\epsilon}$, $M_{\eta}$ and $M_{\epsilon\eta}$, from the ALL data set. 
  For bounded parameters, shown are mean values and 68\% credible intervals.
  For unbounded parameters, 95\% credible intervals are only shown with parentheses.
  }
  \label{tbl:HSR_constr}
  \end{center}
\end{table}

\section{Summary and future outlook} \label{sec:summary}
We have investigated constraints on the primordial power spectra 
and comparison of single-field slow-roll inflation models using 
data from recent observations of CMB combined with measurements of 
BAO, SN and $H_0$. 
By employing Bayesian model selection, we found that 
a model with the scale-invariant HZ spectrum is strongly 
disfavored from current data, in comparison with several 
simple models allowing scale-dependence of the power spectra.
We have also proposed an optimal constraint on the primordial 
power spectra of Eqs.~(\ref{eq:opt_A_s}-\ref{eq:opt_n_s}).

Adopting our somewhat artificial 
division of models for single-field slow-roll inflation, 
$M_\epsilon$, $M_\eta$ and $M_{\epsilon\eta}$, 
with a theoretical prior $N>25$, we have found the Bayes evidences
for these models from current data are almost comparable, 
but there is a slight but consistent preference for 
$M_\epsilon$ over the others from various data sets.
We have demonstrated several extension in order to 
identify the origin of the preference for $M_\epsilon$. 
A presence of the tensor perturbation is not signified 
and we have found that the preference originates from a prior 
$N>25$ which appears to make a prior ranges for 
$\epsilon_*$ center around the region of high likelihood. 
We have also found that higher order slow-roll parameters are not
required from current data. Thus we conclude that while 
simple models of single-field slow-roll inflation can describe 
current cosmological observations, data is not enough 
constraining to distinguish these models we adopted.

Planck~\cite{:2006uk} and several other ground-based CMB surveys, such as 
QUIET~\cite{Samtleben:2008rb}, PolarBeaR~\cite{PolarBear}, etc. are now under way. 
In the near future, the tensor perturbation will be probed down to $r\simeq0.01$ by these surveys.
This allow us to discriminate models of single-field inflation through the
prediction on the primordial power spectra or the slow-roll parameters. Indeed, one of our division
of models, $M_\epsilon$, will be strongly refused, if the primordial tensor perturbation is not 
detected by those surveys. Given data from these future surveys, employment of Bayesian 
model selection would be of great help in several ways, such as constraining cosmological 
parameters in an optimal way, and assessing a support for a cosmological model statistically.
\bigskip
\bigskip

\noindent 
\section*{Acknowledgment}
T.S. would like to thank the Japan Society for the Promotion of
Science for financial support. This work is supported by Grant-in-Aid for Scientific research from the Ministry of 
Education, Science, Sports, and Culture (MEXT), Japan, under Contract No. 14102004 (M.K.),
and also by World Premier International Research Center Initiative, MEXT, Japan.



\begin{thebibliography}{}

\bibitem{Starobinsky:1980te}
  A.~A.~Starobinsky,
  Phys.\ Lett.\  B {\bf 91} (1980) 99.

\bibitem{Guth:1980zm}
  A.~H.~Guth,
  Phys.\ Rev.\  D {\bf 23}, 347 (1981).

\bibitem{Sato:1980yn}
  K.~Sato,
  Mon.\ Not.\ Roy.\ Astron.\ Soc.\  {\bf 195}, 467 (1981).

\bibitem{Linde:1981mu}
  A.~D.~Linde,
  Phys.\ Lett.\  B {\bf 108}, 389 (1982).

\bibitem{Albrecht:1982wi}
  A.~J.~Albrecht and P.~J.~Steinhardt,
  Phys.\ Rev.\ Lett.\  {\bf 48}, 1220 (1982).

\bibitem{Starobinsky:1979ty}
  A.~A.~Starobinsky,
  JETP Lett.\  {\bf 30} (1979) 682
  [Pisma Zh.\ Eksp.\ Teor.\ Fiz.\  {\bf 30} (1979) 719].

\bibitem{Mukhanov:1981xt}
  V.~F.~Mukhanov and G.~V.~Chibisov,
  JETP Lett.\  {\bf 33} (1981) 532
  [Pisma Zh.\ Eksp.\ Teor.\ Fiz.\  {\bf 33} (1981) 549].

\bibitem{Hawking:1982cz}
  S.~W.~Hawking,
  Phys.\ Lett.\  B {\bf 115}, 295 (1982).
  
\bibitem{Starobinsky:1982ee}
  A.~A.~Starobinsky,
  Phys.\ Lett.\  B {\bf 117} (1982) 175.

\bibitem{Guth:1982ec}
  A.~H.~Guth and S.~Y.~Pi,
  Phys.\ Rev.\ Lett.\  {\bf 49}, 1110 (1982).

\bibitem{Linde:1982uu}
  A.~D.~Linde,
  Phys.\ Lett.\  B {\bf 116}, 335 (1982).
  
\bibitem{Bardeen:1983qw}
  J.~M.~Bardeen, P.~J.~Steinhardt and M.~S.~Turner,
  Phys.\ Rev.\  D {\bf 28}, 679 (1983).

\bibitem{Abbott:1984fp}
  L.~F.~Abbott and M.~B.~Wise,
  Nucl.\ Phys.\  B {\bf 244}, 541 (1984).

\bibitem{Copeland:1993jj}
  E.~J.~Copeland, E.~W.~Kolb, A.~R.~Liddle and J.~E.~Lidsey,
  Phys.\ Rev.\  D {\bf 48}, 2529 (1993)
  [arXiv:hep-ph/9303288].

\bibitem{Lidsey:1995np}
  J.~E.~Lidsey, A.~R.~Liddle, E.~W.~Kolb, E.~J.~Copeland, T.~Barreiro and M.~Abney,
  Rev.\ Mod.\ Phys.\  {\bf 69}, 373 (1997)
  [arXiv:astro-ph/9508078].

\bibitem{Leach:2002dw}
  S.~M.~Leach and A.~R.~Liddle,
  Mon.\ Not.\ Roy.\ Astron.\ Soc.\  {\bf 341}, 1151 (2003)
  [arXiv:astro-ph/0207213].

\bibitem{Easther:2002rw}
  R.~Easther and W.~H.~Kinney,
  Phys.\ Rev.\  D {\bf 67}, 043511 (2003)
  [arXiv:astro-ph/0210345].

\bibitem{Kinney:2003uw}
  W.~H.~Kinney, E.~W.~Kolb, A.~Melchiorri and A.~Riotto,
  Phys.\ Rev.\  D {\bf 69}, 103516 (2004)
  [arXiv:hep-ph/0305130].
  
\bibitem{Leach:2003us}
  S.~M.~Leach and A.~R.~Liddle,
  Phys.\ Rev.\  D {\bf 68}, 123508 (2003)
  [arXiv:astro-ph/0306305].

\bibitem{Peiris:2006ug}
  H.~Peiris and R.~Easther,
  JCAP {\bf 0607}, 002 (2006)
  [arXiv:astro-ph/0603587].

\bibitem{Kinney:2006qm}
  W.~H.~Kinney, E.~W.~Kolb, A.~Melchiorri and A.~Riotto,
  Phys.\ Rev.\  D {\bf 74}, 023502 (2006)
  [arXiv:astro-ph/0605338].

\bibitem{Martin:2006rs}
  J.~Martin and C.~Ringeval,
  JCAP {\bf 0608}, 009 (2006)
  [arXiv:astro-ph/0605367].

\bibitem{Peiris:2006sj}
  H.~Peiris and R.~Easther,
  JCAP {\bf 0610}, 017 (2006)
  [arXiv:astro-ph/0609003].

\bibitem{Finelli:2006fi}
  F.~Finelli, M.~Rianna and N.~Mandolesi,
  JCAP {\bf 0612}, 006 (2006)
  [arXiv:astro-ph/0608277].

\bibitem{Lesgourgues:2007gp}
  J.~Lesgourgues and W.~Valkenburg,
  Phys.\ Rev.\  D {\bf 75}, 123519 (2007)
  [arXiv:astro-ph/0703625].

\bibitem{Powell:2007gu}
  B.~A.~Powell and W.~H.~Kinney,
  JCAP {\bf 0708}, 006 (2007)
  [arXiv:0706.1982 [astro-ph]].

\bibitem{Lesgourgues:2007aa}
  J.~Lesgourgues, A.~A.~Starobinsky and W.~Valkenburg,
  JCAP {\bf 0801}, 010 (2008)
  [arXiv:0710.1630 [astro-ph]].

\bibitem{Bean:2008ga}
  R.~Bean, D.~J.~H.~Chung and G.~Geshnizjani,
  Phys.\ Rev.\  D {\bf 78}, 023517 (2008)
  [arXiv:0801.0742 [astro-ph]].

\bibitem{Hamann:2008pb}
  J.~Hamann, J.~Lesgourgues and W.~Valkenburg,
  JCAP {\bf 0804}, 016 (2008)
  [arXiv:0802.0505 [astro-ph]].

\bibitem{Adshead:2008vn}
  P.~Adshead and R.~Easther,
  JCAP {\bf 0810}, 047 (2008)
  [arXiv:0802.3898 [astro-ph]].
  
\bibitem{Agarwal:2008ah}
  N.~Agarwal and R.~Bean,
  Phys.\ Rev.\  D {\bf 79}, 023503 (2009)
  [arXiv:0809.2798 [astro-ph]].
  
\bibitem{Powell:2008bi}
  B.~A.~Powell, K.~Tzirakis and W.~H.~Kinney,
  JCAP {\bf 0904}, 019 (2009)
  [arXiv:0812.1797 [astro-ph]].
    

\bibitem{Komatsu:2008hk}
  E.~Komatsu {\it et al.}  [WMAP Collaboration],
  Astrophys.\ J.\ Suppl.\  {\bf 180}, 330 (2009)
  [arXiv:0803.0547 [astro-ph]].


\bibitem{Slosar:2002dc}
  A.~Slosar {\it et al.},
  Mon.\ Not.\ Roy.\ Astron.\ Soc.\  {\bf 341}, L29 (2003)
  [arXiv:astro-ph/0212497].

\bibitem{Beltran:2005xd}
  M.~Beltran, J.~Garcia-Bellido, J.~Lesgourgues, A.~R.~Liddle and A.~Slosar,
  Phys.\ Rev.\  D {\bf 71}, 063532 (2005)
  [arXiv:astro-ph/0501477].
  
\bibitem{Trotta:2005ar}
  R.~Trotta,
  Mon.\ Not.\ Roy.\ Astron.\ Soc.\  {\bf 378}, 72 (2007)
  [arXiv:astro-ph/0504022].

\bibitem{Bridges:2005br}
  M.~Bridges, A.~N.~Lasenby and M.~P.~Hobson,
  Mon.\ Not.\ Roy.\ Astron.\ Soc.\  {\bf 369}, 1123 (2006)
  [arXiv:astro-ph/0511573].

\bibitem{Kunz:2006mc}
  M.~Kunz, R.~Trotta and D.~Parkinson,
  Phys.\ Rev.\  D {\bf 74}, 023503 (2006)
  [arXiv:astro-ph/0602378].
  
\bibitem{Parkinson:2006ku}
  D.~Parkinson, P.~Mukherjee and A.~R.~Liddle,
  Phys.\ Rev.\  D {\bf 73}, 123523 (2006)
  [arXiv:astro-ph/0605003].

\bibitem{Bridges:2006zm}
  M.~Bridges, A.~N.~Lasenby and M.~P.~Hobson,
  arXiv:astro-ph/0607404.
  
\bibitem{Pahud:2006kv}
  C.~Pahud, A.~R.~Liddle, P.~Mukherjee and D.~Parkinson,
  Phys.\ Rev.\  D {\bf 73}, 123524 (2006)
  [arXiv:astro-ph/0605004].

\bibitem{Liddle:2006tc}
  A.~R.~Liddle, P.~Mukherjee and D.~Parkinson,
  arXiv:astro-ph/0608184.
  
\bibitem{Heavens:2007ka}
  A.~F.~Heavens, T.~D.~Kitching and L.~Verde,
  Mon.\ Not.\ Roy.\ Astron.\ Soc.\  {\bf 380}, 1029 (2007)
  [arXiv:astro-ph/0703191].

\bibitem{Gordon:2007xm}
  C.~Gordon and R.~Trotta,
  Mon.\ Not.\ Roy.\ Astron.\ Soc.\  {\bf 382}, 1859 (2007)
  [arXiv:0706.3014 [astro-ph]].

\bibitem{Mukherjee:2008pq}
  P.~Mukherjee and A.~R.~Liddle,
  arXiv:0803.1738 [astro-ph].

\bibitem{Trotta:2008qt}
  R.~Trotta,
  Contemp.\ Phys.\  {\bf 49}, 71 (2008)
  [arXiv:0803.4089 [astro-ph]].
  
\bibitem{Mukherjee:2008zzb}
  P.~Mukherjee and D.~Parkinson,
  Int.\ J.\ Mod.\ Phys.\  A {\bf 23} (2008) 787.

\bibitem{Bridges:2008ta}
  M.~Bridges, F.~Feroz, M.~P.~Hobson and A.~N.~Lasenby,
  arXiv:0812.3541 [astro-ph].

\bibitem{Liddle:2009xe}
  A.~R.~Liddle,
  arXiv:0903.4210 [hep-th].
  
\bibitem{Sollom:2009vd}
  I.~Sollom, A.~Challinor and M.~P.~Hobson,
  Phys.\ Rev.\  D {\bf 79}, 123521 (2009)
  [arXiv:0903.5257 [astro-ph.CO]].

\bibitem{Ichikawa:2009ir}
  K.~Ichikawa, M.~Kawasaki, K.~Nakayama, T.~Sekiguchi and T.~Takahashi,
  JCAP {\bf 0908}, 013 (2009)
  [arXiv:0905.2237 [astro-ph.CO]].
  
\bibitem{Valiviita:2009bp}
  J.~Valiviita and T.~Giannantonio,
  arXiv:0909.5190 [astro-ph.CO].

\bibitem{Ballesteros:2007te}
  G.~Ballesteros, J.~A.~Casas, J.~R.~Espinosa, R.~Ruiz de Austri and R.~Trotta,
  JCAP {\bf 0803}, 018 (2008)
  [arXiv:0711.3436 [hep-ph]].

\bibitem{Baumann:2008aq}
  D.~Baumann {\it et al.}  [CMBPol Study Team Collaboration],
  AIP Conf.\ Proc.\  {\bf 1141}, 10 (2009)
  [arXiv:0811.3919 [astro-ph]].



\bibitem{Leach:2002ar}
  S.~M.~Leach, A.~R.~Liddle, J.~Martin and D.~J.~Schwarz,
  Phys.\ Rev.\  D {\bf 66}, 023515 (2002)
  [arXiv:astro-ph/0202094].
  
\bibitem{Hoffman:2000ue}
  M.~B.~Hoffman and M.~S.~Turner,
  Phys.\ Rev.\  D {\bf 64}, 023506 (2001)
  [arXiv:astro-ph/0006321].

\bibitem{Kinney:2002qn}
  W.~H.~Kinney,
  Phys.\ Rev.\  D {\bf 66}, 083508 (2002)
  [arXiv:astro-ph/0206032].
  

\bibitem{Liddle:2003py}
  A.~R.~Liddle,
  Phys.\ Rev.\  D {\bf 68}, 103504 (2003)
  [arXiv:astro-ph/0307286].
  
\bibitem{Stewart:1993bc}
  E.~D.~Stewart and D.~H.~Lyth,
  Phys.\ Lett.\  B {\bf 302}, 171 (1993)
  [arXiv:gr-qc/9302019].

\bibitem{Peiris:2008be}
  H.~V.~Peiris and R.~Easther,
  JCAP {\bf 0807}, 024 (2008)
  [arXiv:0805.2154 [astro-ph]].

\bibitem{Cortes:2007ak}
  M.~Cortes, A.~R.~Liddle and P.~Mukherjee,
  Phys.\ Rev.\  D {\bf 75}, 083520 (2007)
  [arXiv:astro-ph/0702170].

\bibitem{Kosowsky:2002zt}
  A.~Kosowsky, M.~Milosavljevic and R.~Jimenez,
  Phys.\ Rev.\  D {\bf 66}, 063007 (2002)
  [arXiv:astro-ph/0206014].

\bibitem{Komatsu:2002wc}
  E.~Komatsu and U.~Seljak,
  Mon.\ Not.\ Roy.\ Astron.\ Soc.\  {\bf 336}, 1256 (2002)
  [arXiv:astro-ph/0205468].
  

\bibitem{Dunkley:2008ie}
  J.~Dunkley {\it et al.}  [WMAP Collaboration],
  Astrophys.\ J.\ Suppl.\  {\bf 180}, 306 (2009)
  [arXiv:0803.0586 [astro-ph]].

\bibitem{Nolta:2008ih}
  M.~R.~Nolta {\it et al.}  [WMAP Collaboration],
  Astrophys.\ J.\ Suppl.\  {\bf 180}, 296 (2009)
  [arXiv:0803.0593 [astro-ph]].

\bibitem{Hinshaw:2008kr}
  G.~Hinshaw {\it et al.}  [WMAP Collaboration],
  Astrophys.\ J.\ Suppl.\  {\bf 180}, 225 (2009)
  [arXiv:0803.0732 [astro-ph]].
  
\bibitem{Reichardt:2008ay}
  C.~L.~Reichardt {\it et al.},
  arXiv:0801.1491 [astro-ph].

\bibitem{Sievers:2005gj}
  J.~L.~Sievers {\it et al.},
  Astrophys.\ J.\  {\bf 660}, 976 (2007)
  [arXiv:astro-ph/0509203].
  
\bibitem{Jones:2005yb}
  W.~C.~Jones {\it et al.},
  Astrophys.\ J.\  {\bf 647}, 823 (2006)
  [arXiv:astro-ph/0507494].

\bibitem{Piacentini:2005yq}
  F.~Piacentini {\it et al.},
  Astrophys.\ J.\  {\bf 647}, 833 (2006)
  [arXiv:astro-ph/0507507].

\bibitem{Montroy:2005yx}
  T.~E.~Montroy {\it et al.},
  Astrophys.\ J.\  {\bf 647}, 813 (2006)
  [arXiv:astro-ph/0507514].

\bibitem{Friedman:2009dt}
  R.~B.~Friedman {\it et al.}  [QUaD collaboration],
  arXiv:0901.4334 [astro-ph.CO].

\bibitem{Reid:2009xm}
  B.~A.~Reid {\it et al.},
  arXiv:0907.1659 [astro-ph.CO].

\bibitem{Finelli:2009bs}
  F.~Finelli, J.~Hamann, S.~M.~Leach and J.~Lesgourgues,
  arXiv:0912.0522 [astro-ph.CO].

\bibitem{Peiris:2009wp}
  H.~V.~Peiris and L.~Verde,
  arXiv:0912.0268 [astro-ph.CO].

\bibitem{Kowalski:2008ez}
  M.~Kowalski {\it et al.}  [Supernova Cosmology Project Collaboration],
  Astrophys.\ J.\  {\bf 686}, 749 (2008)
  [arXiv:0804.4142 [astro-ph]].

\bibitem{Percival:2009xn}
  W.~J.~Percival {\it et al.},
  arXiv:0907.1660 [astro-ph.CO].

\bibitem{Riess:2009pu}
  A.~G.~Riess {\it et al.},
  arXiv:0905.0695 [astro-ph.CO].


\bibitem{Feroz:2008xx}
  F.~Feroz, M.~P.~Hobson and M.~Bridges,
  arXiv:0809.3437 [astro-ph].

\bibitem{Lewis:2002ah}
  A.~Lewis and S.~Bridle,
  Phys.\ Rev.\  D {\bf 66}, 103511 (2002)
  [arXiv:astro-ph/0205436].
  
\bibitem{Skilling:2004}
Skilling J., 2004, AIP Conference Proceedings of the 24th International 
Workshop on Bayesian Inference and Maximum Entropy Methods in Science and 
Engineering, Vol. 735, pp. 395-405

\bibitem{Mukherjee:2005wg}
  P.~Mukherjee, D.~Parkinson and A.~R.~Liddle,
  Astrophys.\ J.\  {\bf 638}, L51 (2006)
  [arXiv:astro-ph/0508461].


\bibitem{:2006uk}
    [Planck Collaboration],
  arXiv:astro-ph/0604069.
  
\bibitem{Samtleben:2008rb}
  D.~Samtleben and f.~t.~Q.~Collaboration,
  Nuovo Cim.\  {\bf 122B}, 1353 (2007)
  [arXiv:0802.2657 [astro-ph]].
  
  \bibitem{PolarBear}
  http://bolo.berkeley.edu/polarbear/
  
  \end{thebibliography}
\end{document}